\documentstyle[twoside,fleqn,espcrc2,epsf]{article}

\newcommand{\be}{\begin{equation}}
\newcommand{\ee}{\end{equation}}
\newcommand{\bea}{\begin{eqnarray}}
\newcommand{\eea}{\end{eqnarray}}

\newcommand{\AmS}{{\protect\the\textfont2
A\kern-.1667em\lower.5ex\hbox{M}\kern-.125emS}}

\title{
\hbox{}
{\small DECEMBER 1994} \hfill {\small WUB 94--37}     \break
                                                              \break
Color flux profiles in SU(2) lattice gauge theory\thanks{
Talk given at Lattice 94 conference, Bielefeld, Germany,
       Sept.\ 27 -- Oct.\ 1, 1994.
We thank the
HLRZ for computer time on the CM-5 at GMD.
Work supported by DFG grants Schi 257/1-4, Schi 257/3-2 and Mu 810/3,
EC project SC1*-CT91-0642 and EC contract CHRX-CT92-0051.}}

\author{\underline{C.~Schlichter$^{\rm a}$},
G.S.~Bali\address{Fachbereich Physik, Bergische Universit\"at,
D-42097 Wuppertal, Germany} and K.~Schilling\address{HLRZ c/o
Forschungszentrum J\"ulich, D-52425 J\"ulich,
Germany}$^{\rm \!\!,a}$}

\begin{document}

\begin{abstract}
Results of a high statistics study of chromo field distributions
around static sources in pure SU(2) gauge theory on lattices of volumes
$16^4$, $32^4$, and $48^3\times 64$ at $\beta=2.5$, $2.635$, and
$2.74$ are presented. We establish string formation up to
physical distances as large as 2 fm.
\end{abstract}

\maketitle

\section{INTRODUCTION}
Various effective models \cite{luescher,effective} originated from the
type II superconductivity scenario of
confinement~\cite{thooft}. Lattice gauge theory, in principle, offers the
laboratory to test such ideas, as it allows for {\em ab initio}
studies from the QCD Lagrangian.  Quenched calculations have reached the
accuracy necessary for quantitative studies of the chromo field
distributions. These are not easily accessible, since (a) the energy
density carries dimension $a^{-4}$ and therefore imposes a lower limit
onto the lattice spacing $a$, due to statistical noise, and (b) one is
forced to work on rather large lattice volumes to attain source
separations sufficient for applicability of the string picture.  On
top of this one is faced with the ubiquitous problem of filtering
ground state signals out of an excited state background.

We have carried out a comprehensive investigation of flux tube
profiles between static $q\bar q$ pairs. Exploitation of
state-of-the-art lattice techniques for statistical noise reduction
and ground state enhancement enables us to observe string formation up
to physical distances of 2 fm, corresponding to more than 30 lattice
sites at $\beta =2.635$.

\section{SIMULATION}

We study lattice volumes of $16^4$ and $32^4$ at $\beta=2.5$
($a\approx 0.083$~fm), $48^3\times 64$ at $\beta=2.635$ ($a\approx
0.054$~fm) and $32^4$ at $\beta=2.74$ ($a\approx 0.041$~fm). The scale
has been computed from the string tension value $\sqrt{K}a=$ 440
MeV. A hybrid of Fabricius-Haan
heatbath and overrelaxation has been applied for the gauge
field update.

The central observables in our present investigation, the action and
energy densities in presence of two static quark sources, \be
\left.\begin{array}{l}\epsilon_R({\bf n})\\ \sigma_R({\bf
n})\end{array}\right\} =\frac{1}{2}\left({\bf E}^2({\bf n}) \pm{\bf
B}^2({\bf n})\right)_{|0,R\rangle-|0\rangle} \ee are calculated by
taking the correlation between smeared Wilson loops $\cal W$ and
plaquettes $\Box$,
\begin{equation}
\langle\Box\rangle_{\cal W}=
\frac{\langle {\cal W}\Box\rangle}{\langle
{\cal W}\rangle} -\langle\Box\rangle\quad.
\end{equation}
The flux-tube profiles are probed by varying the spatial position of
the plaquette which acts like a
chromo electric (magnetic) Hall detector. To minimize
contaminations from excited states, the temporal position is chosen as
close as possible to the center of the Wilson loop.
By averaging
two (four) adjacent electric (magnetic) plaquettes,
we arrive at operator insertions
that are symmetric in respect to a given
(spatial) lattice site $\bf n$.

For measurement of the colour field distributions we have restricted
ourselves to on-axis separations of the two sources. All even
distances $R=2,4,\ldots,R_{\max}$ with $R_{\max} = 8,24,36$ for
$L_S=16,32,48$, respectively, have been realized. In order to identify
the asymptotic plateau, $T$ was varied up to $T=6$. The colour field
distributions have been measured up to a transverse distance
$n_{\perp}=2$ along the entire $Q\bar{Q}$ axis. In between the two
sources and up to 2 lattice spacings outside the sources, the
transverse distance was increased to $n_{\perp,\max}=6,10,15$ for the
three lattice extents $L_S=16,32,48$, respectively.  In addition to
``on-axis'' positions, ${\bf n}=(n_1,n_2,0)$, we chose plane-diagonal
points ${\bf n}=(n_1,n_2,n_2)$ with $n_2<n_{\perp,\max}/\sqrt{2}$.  We
averaged over various coordinates ${\bf n}$, exploiting the
cylindrical and reflection symmetry of the problem.\\ The field
measurements have been taken every 100 sweeps at $\beta=2.5,2.74$ and
every 200 sweeps at $\beta=2.635$. At
$(\beta,V)=(2.5,16^4),(2.5,32^4),(2.635,48^3\times64),(2.74,32^4)$,
8680, 2046, 248, 1480 measurements have been taken respectively.

The temporal parts of the Wilson loops, appearing in the colour field
correlator, have been link integrated while the spatial parts have
been smeared. The former procedure is applied to improve the
signal to noise ratio, the latter is essential for getting early (in
$T$) ground state dominance.  A more detailed description of the
lattice methods, analysis procedure and results can be found in
Ref.~\cite{cernth7413}. (A data base of colour images of the
flux tubes and related quantities can be accessed via anonymous ftp
from wpts0.physik.uni-wuppertal.de. The (compressed) .rgb and .ps
files can be found in the directory pub/colorflux.)

\section{RESULTS}
\begin{figure}[htb]
\vskip -.5cm
\unitlength1cm
\begin{picture}(7.5,6)
\put(0,0){\epsfxsize=7.5cm\epsfbox{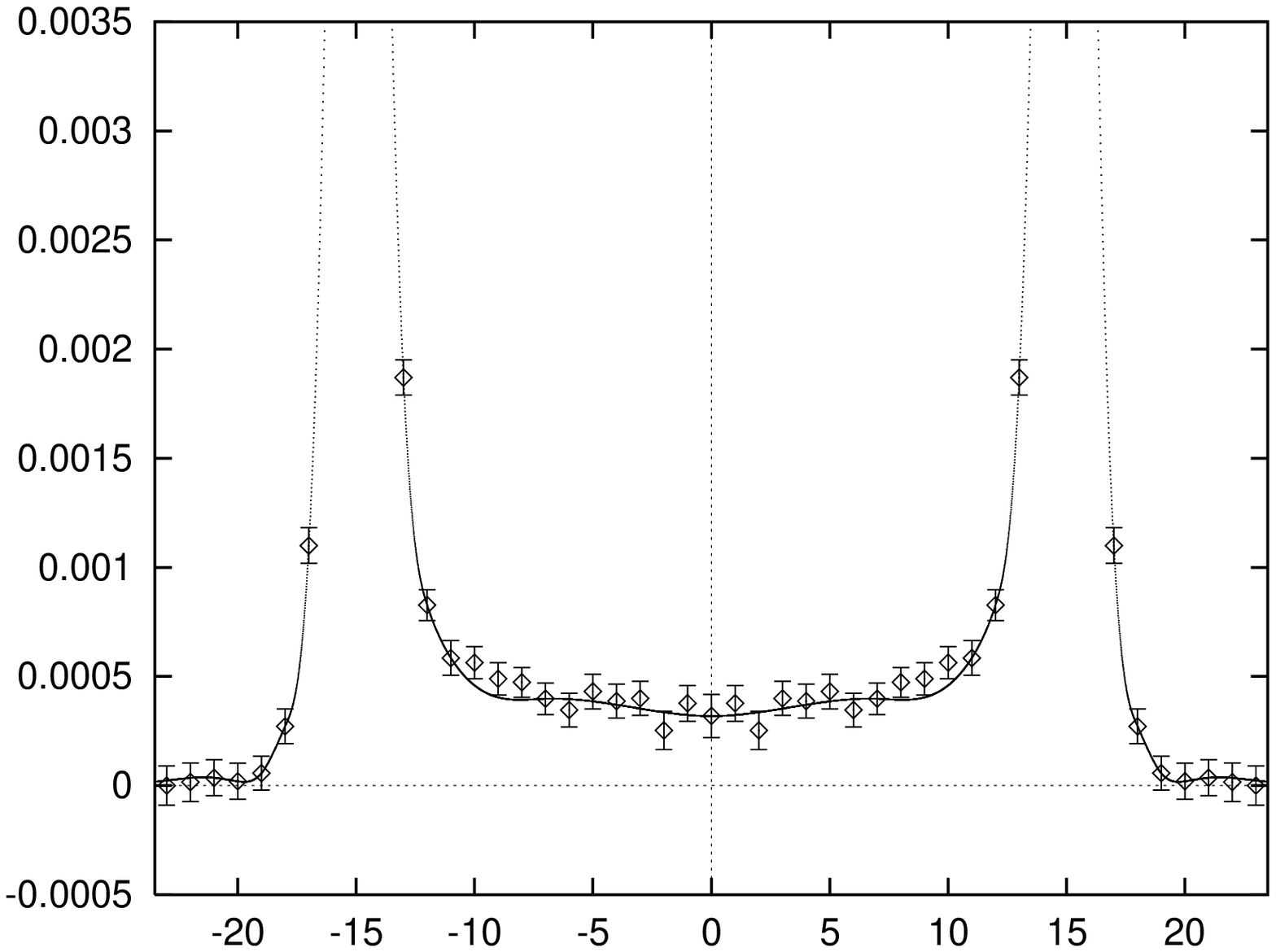}}
\put(-0.3,4.4){\mbox{$\sigma$}}
\put(4,-0.3){\mbox{$n_1$}}
\end{picture}
\vskip -.5cm
\caption{Longitudinal action density profile in lattice units at
$\beta=2.635$, $r=30a\approx 1.6$~fm.}
\label{long}
\vskip -.5cm
\end{figure}

Here, we restrict ourselves to some selected topics.  In
Fig.~\ref{long} the longitudinal action density profile for a quark
separation $r\approx 1.6$ fm is displayed. 
The string, connecting the sources, is (almost) homogenious
within a region of extent 1~fm.

\begin{figure}[htb]
\vskip -.5cm
\unitlength1cm
\begin{picture}(7.5,6)
\put(0,0){\epsfxsize=7.5cm\epsfbox{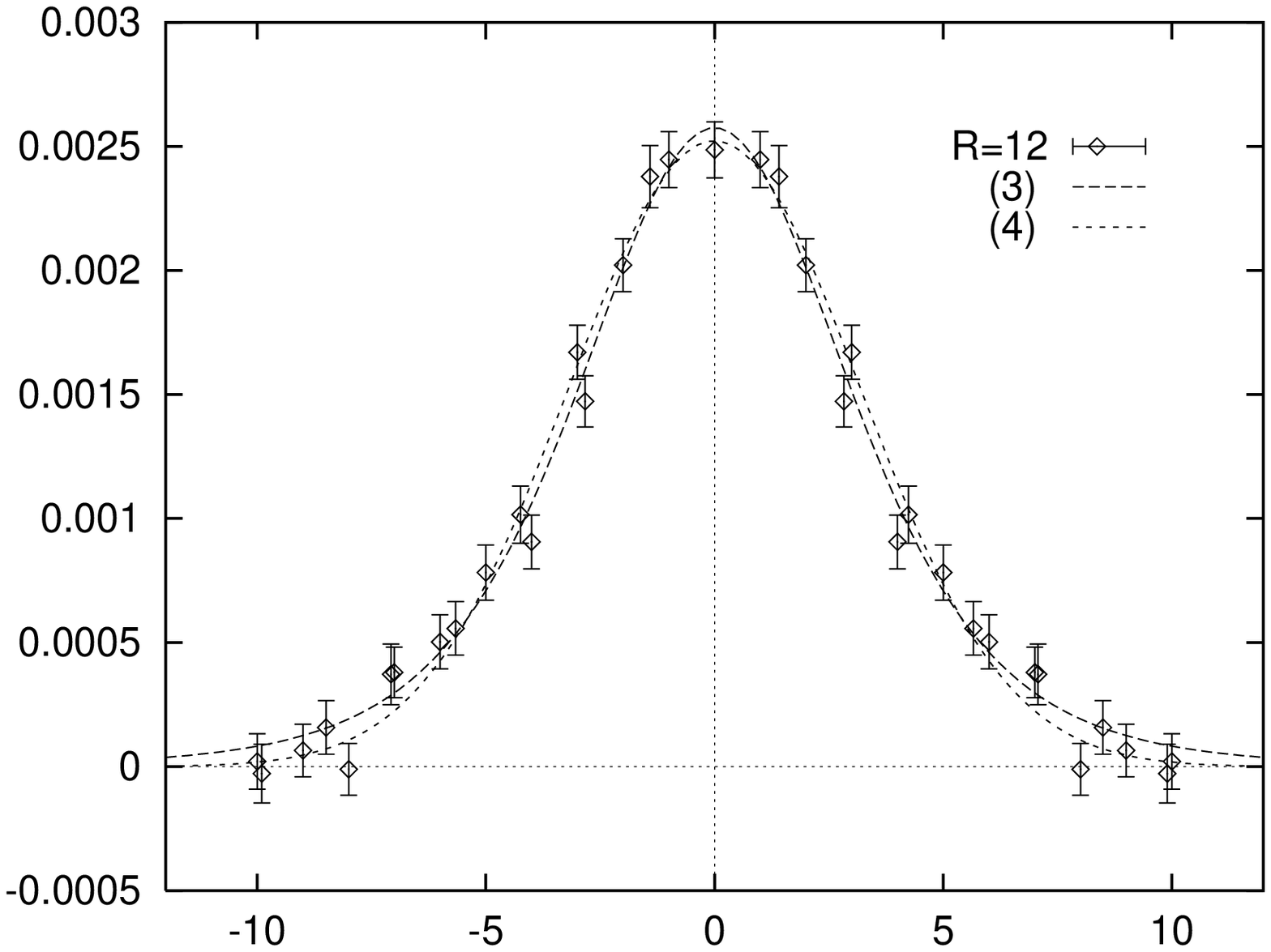}}
\put(-.3,4.4){\mbox{$\sigma$}}
\put(4,-0.3){\mbox{$n_\perp$}}
\end{picture}
\vskip -.5cm
\caption{Central transverse action density profile in lattice
units at $\beta=2.5$,
$r=12a\approx 1.0$~fm, together with dipole (3) and Gaussian (4) fit
curves.}
\label{trans}
\vskip -.5cm
\end{figure}

The field distributions within the center plane of the flux tube have
been studied in detail (see Fig.~\ref{trans}). For small source
separations perturbation theory should apply and one might expect the
distribution to agree with a dipole shape (up to lattice
artefacts):
\begin{equation}
\label{dipole}
f_d(n_{\perp})\propto\frac{1}{(4\pi)}
\frac{4\delta^2}{\left(\delta^2+n_{\perp}^2\right)^3}
\end{equation}
with $\delta=R/2$. $n_{\perp}$ denotes the distance from the
$q\bar q$~axis. 
Fits to the above parametrization have been performed,
in which $\delta$ has been treated as a free parameter.
At large (compared to the width of the flux
tube) source separation, the string picture 
should apply and we expect a Gaussian shape,
at least for small $n_{\perp}$:
\begin{equation}
\label{gauss}
f_g(n_{\perp})\propto\frac{1}{2\pi\delta^4}
\exp\left(-\frac{n_{\perp}^2}{\delta^2}\right)\quad.
\end{equation}

Due to cancellations of electric and magnetic
components, precise data for the energy density have only been
obtained for $r\le 0.75$~fm. Within this region, where the string is
not yet developed the data is well described
by leading order lattice perturbation theory with an effective coupling,
increasing with $R$ in agreement with asymptotic
freedom. The action density data allow for fits up to distances of
$2$~fm. Ans\"atze eqn.\ (\ref{dipole}) and (\ref{gauss}) describe the data
equally well as can be seen from Fig.~\ref{trans}.

\begin{figure}[htb]
\unitlength1cm
\begin{picture}(7.5,6)
\put(0,0){\epsfxsize=7.5cm\epsfbox{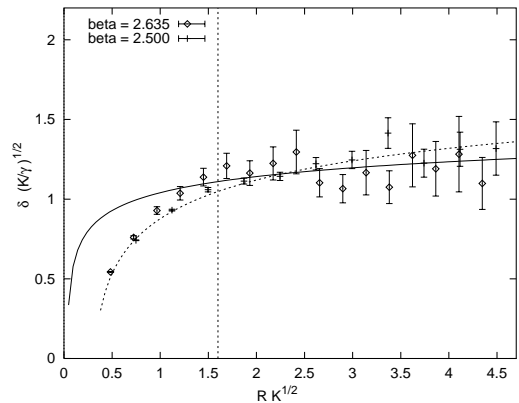}}
\end{picture}
\vskip -.5cm
\caption{Width of the action density flux tube,
$\delta$, against $R$ in units of
the string tension.}
\vskip -.5cm
\label{width}
\end{figure}

The root mean square width, $\delta$, of the flux tube is physically
more interesting than the half width which can be read off from plots
like Fig.~\ref{trans}.  In principle, $\delta$ can be extracted from
fits but strongly depends on the underlying
parametrization. As we are mainly
interested in the $R$ dependence, especially in probing the predicted
logarithmic asymptotic increase of $\delta(R)$~\cite{luescher},
we use a geometric
definition of the width~\cite{cernth7413} that avoids large
errors on ratios of $\delta$s, obtained at different quark
separations. It is based on the assumption that the transverse shape
of the distribution is independent of $R$ which holds true to the
right of the vertical line in Fig.~\ref{width}. The model dependence
of $\delta^2$ has been cast into a (multiplicative)
geometry factor $\gamma$ which is of order one.

The width increases rapidly until it saturates at $r\approx 1$~fm at
a value 0.5~fm$\leq\delta a\leq$ 0.75~fm
where the error is mainly due to the uncertainty in $\gamma$.
Combining the string expectation,
$\delta^2=\delta_0^2\ln(R/R_0)$,
with our data, yields
an inverse cut-off wavelength $R_0^{-1}\approx
30\sqrt{K}$ (full curve) when 
$\delta_0$ is constrained to the
universal value $\delta_0=1/(\pi\sqrt{K})$~\cite{luescher}.
Without restrictions on $\delta_0$ we find
the limits $\delta_0\leq 2.2/(\pi\sqrt{K})$ and $R_0^{-1}\geq
3\sqrt{K}$, respectively (dashed curve).  In case of the simpler case
of 3d $Z_2$ gauge theory, Gliozzi has observed a more pronounced logarithmic
increase of $\delta(R)$~\cite{gliozzi}.

\section{CONCLUSIONS}

We have demonstrated that Wilson loop-plaquette correlations offer a
viable access to a lattice study of the flux tube problem on the required
length scale of one to two fm.  
The crucial ingredient of our method is the smearing of the parallel
transporter within the bilocal $Q\overline{Q}$ creation operator. This
secures a controlled ground state preparation of long flux tubes
within few lattice time slices.

As a result, we can observe flux formation of the action density over
source separations as large as 2~fm on lattices with resolution .05~fm
and extent 2.7~fm.
A logarithmic increase of the width as suggested by string pictures 
for the ``asymptotic'' $R$ region is consistent with
our data, suggesting a surprisingly large ultra violet cut-off on
inverse wavelengths in effective string models, $r_0^{-1}\gg 1$~GeV.

\vskip-1.cm


\begin{thebibliography}{99}
\newcommand{\np}{Nucl.\ Phys.\ } 
\newcommand{\npp}{Nucl.\ Phys.\ [Proc.\ Suppl.]}   
\newcommand{\pl}{Phys.\ Lett.\ }    
\newcommand{\pr}{Phys.\ Rev.\ }
\newcommand{\prl}{Phys.\ Rev.\ Lett.\ }
\bibitem{luescher}{M.\ L\"uscher, G.\ M\"unster, and P.\ Weiss,
\np { B180} (1981) 1.}
\bibitem{effective} {
M.~Baker, J.S.~Ball, and F.~Zachariasen,
Phys.~Rep.\ { 209} (1991) 73;
A.~Chodos et al.\ \pr { D9} (1974) 3471.}
\bibitem{thooft}{
G.\ t'Hooft, in High Energy Physics,
Ed.\ A.\ Zichichi, Edotorice Compositori, Bologna (1975);
S.\ Mandelstam, Phys.~Rep.\ 23C (1976) 245.}
\bibitem{cernth7413}{G.S.~Bali, K.~Schilling, Ch.~Schlichter,
preprint CERN-TH.7413/94 (1994).}
\bibitem{gliozzi}{F.~Gliozzi, this volume.}
\end{thebibliography}
\end{document}